\documentclass[twocolumn,showpacs,superscriptaddress,preprintnumbers,amsmath,amssymb]{revtex4}

\usepackage{graphicx}
\usepackage{dcolumn}
\usepackage{bm}
\usepackage{epsfig}
\usepackage{subfigure}

\begin{document}

\title{A Pseudospectral Method for Optimal Control\\ of Open Quantum Systems}

\author{Jr-Shin Li}
\email{jsli@seas.wustl.edu}
\affiliation{Department of Electrical and Systems Engineering, Washington University in St. Louis, Missouri, USA}
\author{Justin Ruths}
\affiliation{Department of Electrical and Systems Engineering, Washington University in St. Louis, Missouri, USA}
\author{Dionisis Stefanatos}
\affiliation{Prefecture of Kefalonia, Argostoli, Kefalonia 28100, Greece}

\date{\today}

\begin{abstract}
In this paper, we present a unified computational method based on pseudospectral approximations for the design of optimal pulse sequences in open quantum systems. The proposed method transforms the problem of optimal pulse design, which is formulated as a continuous time optimal control problem, to a finite dimensional constrained nonlinear programming problem. This resulting optimization problem can then be solved using existing numerical optimization suites. We apply the Legendre pseudospectral method to a series of optimal control problems on open quantum systems that arise in Nuclear Magnetic Resonance (NMR) spectroscopy in liquids. These problems have been well studied in previous literature and analytical optimal controls have been found. We find an excellent agreement between the maximum transfer efficiency produced by our computational method and the analytical expressions. Moreover, our method permits us to extend the analysis and address practical concerns, including smoothing discontinuous controls as well as deriving minimum energy controls. The method is not restricted to the systems studied in this article but is universal to every open quantum system whose performance is limited by dissipation.
\end{abstract}

\maketitle

\section{Introduction}
The problem of relaxation is ubiquitous in all applications involving coherent control of quantum mechanical phenomena. In these applications, the quantum system of interest interacts with its environment (open quantum system) and relaxes back to some equilibrium state \cite{Breuer07}. This relaxation effect leads to degraded signal recovery and, in turn, to the loss of experimental information. Optimal manipulation of open quantum systems in such a way as to produce desired evolutions while minimizing relaxation losses has been a long-standing and challenging problem in the area of quantum control.

Various methods employing optimization techniques and principles of optimal control have been developed for the design of pulse sequences that can be used to manipulate quantum systems in an optimal manner. However, the large majority of them is limited to deal with closed quantum systems \cite{Kobzar_ExploringI_2004,Kobzar_ExploringII_2008,Pauly,Conolly,Mao,Navin_TO,Pryor}. Recently, relaxation-optimized pulse sequences that maximize the performance of open quantum systems have emerged. In some simple cases, for example maximizing polarization transfer between a pair of coupled spins in the presence of relaxation, the optimal pulses have been derived analytically using optimal control theory \cite{Khaneja03_1,Khaneja03_2,Stefanatos04}. For more general cases, gradient ascent algorithms were proposed to optimize pulse sequences for optimally steering the dynamics of coupled nuclear spins \cite{Navin_GA,Ohtsuki,Skinner1,BBCROP,TROPIC,Stefanatos05}. These algorithms, while successful, rely heavily on the computation of an analytic expression for the system evolution propagator and gradients as well as a large number of discretizations over which to evolve the system. This results in expensive computational power, and gradient ascent algorithms in general inherit slow linear convergence rates \cite{Bertsekas}.

In this article, we present a unified computational method for optimal pulse sequence design based on pseudospectral approximations. This paper is organized as follows. In the next section, we formulate optimal control problems in open quantum systems and introduce pseudospectral methods. In Section \ref{sec:example}, we present several examples to demonstrate the robustness of pseudospectral methods for optimal pulse sequence design. The systems in the examples have been thoroughly studied in our previous work or by others.

\section{Pseudospectral Methods for Open Quantum Systems}
\label{sec:ps}
For an open quantum system, the evolution of its density matrix is not unitary. In many applications of interest, the environment can be approximated as an infinite thermostat the state of which never changes. Under this assumption, so-called the Markovian approximation, it is possible to write the evolution of the density matrix of an open system (master equation) alone in the Lindblad form \cite{Lindblad76}
\begin{equation}
\label{ro} \dot{\rho}= -i[H(t),\rho]-L(\rho)\ , \qquad (\hbar=1)\;,
\end{equation}
where $H(t)$ is the system Hamiltonian that generates unitary evolution while the term $L(\rho)$ models relaxation (nonunitary dynamics). The general form of $L$ is
\begin{equation}
\label{Lind} L(\cdot)= \sum_{\alpha,\,\beta}
k_{\alpha\beta}[V_{\alpha}, [V^{\dag}_{\beta}, .\,]]\;,
\end{equation}
where $V_{\alpha,\,\beta}$ are operators that represent various relaxation mechanisms and $k_{\alpha\beta}$ are coefficients that depend on the physical parameters of the system. The Hamiltonian $H(t)$ has the general form
\begin{equation}
\label{unitary} H(t)= H_f+\sum_{i=1}^{m}u_i(t)H_i \;,
\end{equation}
where $H_f$ is the free evolution Hamiltonian and $\sum_{i=1}^{m}u_i(t)H_i$ is the so-called control Hamiltonian. The later is used to manipulate the open quantum system by application of electromagnetic pulses $u_i(t)$ of appropriate shape and frequency.

A typical problem of controlling an (finite-dimensional) open quantum system has the following form: starting from some initial state $\rho(0)$ at $t=0$, find optimal pulses $u_i(t)$, $0\leq t \leq T$, that bring the final density matrix $\rho(T)$ at $t=T$ as ``close'' as possible to some target operator $O$. More precisely, find $u_i(t)$ that maximize the final expectation value of $O$, $\langle O\rangle (T)={\rm trace}\{\rho(T)O\}$. Using the master equation (\ref{ro}) and the relation $d\langle O\rangle/dt=\mbox{trace} \{ \dot{\rho} O\}$ that holds for a time-independent operator, we can find a system of ordinary differential equations that describe the time evolution of an open system for the desired transfer $\rho(0)\rightarrow O$ \cite{Goldman88}
\begin{equation}
\label{bilinear}
\dot{x}=\Big[\mathcal{H}_f+\sum_{i=1}^{m}u_i(t)\mathcal{H}_i\Big]x\;,
\end{equation}
where $x=(x_1,\ldots,x_n)^T\in \mathbb{R}^{\rm n}$ is the state vector whose elements are expectation values of the operators participating in the transfer (for example usually $x_n(t)=\langle O\rangle(t)$), $\mathcal{H}_f, \mathcal{H}_i\in\mathbb{R}^{\rm n\times n}$ are square matrices corresponding to operators $H_f, H_i$ under a fixed basis of the state space (Hilbert space) and $t\in[0,T]$. This gives rise to an optimal control problem that starting from an initial state $x(0)$ (which is related to $\rho(0)$), find the controls $u_i(t)$, $t\in[0,T]$, that maximize $x_n(T)=\langle O\rangle(T)$ subject to the system evolution equations as in (\ref{bilinear}). Specific examples are given in the next section.

Practical considerations such as power and time constraints guide us to consider a more general cost function (the quantity that we want to maximize or minimize)
\begin{equation}
\label{cost}
\mbox{min}\;\;\varphi(T,x(T))+\int_0^T \mathcal{L}(x(t),u(t))dt\;,
\end{equation}
where $u=(u_1, u_2,\ldots, u_m)^T$ is the control vector, $\varphi$ is the terminal cost depending on the final state at the terminal time $t=T$, and $\mathcal{L}$ is the running cost depending on the time history of the state and control variables, $x$ and $u$. For example, if $\varphi=-x_n(T)$ and $\mathcal{L}=0$, then (\ref{cost}) is equivalent to maximizing $x_n(T)$ as mentioned above, while if $\varphi=0$ and $\mathcal{L}=\sum_{i=1}^{m}u_i^2$, then (\ref{cost}) is equivalent to minimizing the energy of the pulses. In many cases of application, not only the initial state $x(0)$ can be specified but also other endpoint constraints may be imposed. They can be expressed in a compact form as
\begin{equation}
\label{endpoint}
e(x(0),x(T))=0.
\end{equation}
Additionally, constraints on the state and control variables satisfied along the path of the system may be imposed, such as the amplitude constraints where $|u_i(t)| \leq M$, for all $t\in [0,T]$, where $M$ is the maximum amplitude of the pulses. Such constraints can be expressed as
\begin{equation}
\label{statecontrol}
g(x(t),u(t))\leq 0.
\end{equation}
This class of optimal control problems of bilinear systems (linear in both state and control) described in \eqref{bilinear} are in general analytically intractable. However, they can be efficiently solved by pseudospectral methods.

Spectral methods involve the expansion of functions in terms of orthogonal polynomial basis functions on the domain $[-1,1]$. Using such a basis leads to \emph{spectral accuracy}, namely, the $k^{\text{th}}$ coefficient of the expansion decays faster than any inverse power of $k$ \cite{canuto_spectral_2006}, which is analogous to the Fourier series expansion for periodic functions. This property of rapid decay from spectral methods is adapted to solve optimal control problems like the one described above. It permits the use of relatively low order polynomials to approximate the control and state trajectory functions, $u(t)$ and $x(t)$.

Since the support of the orthogonal polynomial bases is on the interval $[-1,1]$, we first transform the optimal control problem from the time domain $t\in[0,T]$ to $\tau\in[-1,1]$ using the simple affine transformation,
\begin{equation*}\label{eq:PSaffine}
 \tau(t)=\frac{2t-T}{T}.
\end{equation*}
\noindent In a redundant use of notation, we make this transition and reuse the same time variable $t$. The transformed optimal control problem can now be written as,
\begin{align}
\min\ \ & \varphi(1,x(1))+\frac{T}{2}\int_{-1}^1 \mathcal{L}(x(t),u(t))dt \nonumber\\
\label{eq:opc-11}
{\rm s.t.}\ \ & \dot{x}=\frac{T}{2}\Big[\mathcal{H}_f+\sum_{i=1}^{m}u_i(t)\mathcal{H}_i\Big]x\\
& e(x(-1),x(1)) = 0\nonumber\\
& g(x(t),u(t))\leq 0.\nonumber
\end{align}
Pseudospectral methods were developed to solve partial differential equations and recently adapted to solve optimal control problems \cite{elnagar_pseudospectral_1995, ross_legendre_2004, wei_kang_convergence_2006, fahroo_costate_2001, williams_gauss--lobatto_2006}. Pseudospectral approximations are a spectral collocation (or interpolation) method in which the differential equation describing the state dynamics is enforced at specific nodes. Spectral collocation is motivated by the Chebyshev Equioscillation Theorem \cite{davis_interpolation_1963} which states that the best $N^{\text{th}}$ order approximating polynomial $p_N^*(f)$ to a continuous function $f$ on the interval $[-1,1]$ is an interpolating polynomial, as evaluated by the uniform norm,
\begin{equation}\label{eq:PSinfnorm}
	\|f-p_N^*(f)\|_\infty = \min_{p\,\in\mathbb{P}_{\rm N}} \|f-p\|_\infty,
\end{equation}
where $\mathbb{P}_{\rm N}$ is the space of all polynomials of degree at most N. Since any $N^{\text{th}}$ order interpolating polynomial can be represented in terms of the Lagrange basis functions (or Lagrange polynomials), we use these functions to express the interpolating approximations of the continuous state and control functions, $x(t)$ and $u(t)$, as in the model \eqref{eq:opc-11}. Given a grid of $N+1$ interpolation nodes within $[-1,1]$, $\Gamma = \{t_0 < t_1 < \dots < t_N\}$, the Lagrange polynomials $\{\ell_k\}\in\mathbb{P_{\rm N}}$, $k\in\{0,1,\ldots,N\}$, are constructed by
\begin{equation*}
    \ell_k(t) = \prod_{i=0 \atop i\ne k}^N \displaystyle\frac{(t-t_i)}{(t_k-t_i)},
\end{equation*}
which are characterized by the $k^{\text{th}}$ polynomial taking unit value at the $k^{\text{th}}$ node of the grid and zero value at all other nodes of the grid, i.e., $\ell_k(t_i) = \delta_{ki}$, where $\delta_{ki}$ is the Kronecker delta function \cite{szego_orthogonal_1959}. Note that Lagrange polynomials form an orthogonal basis with respect to the discrete inner product $\langle p,q \rangle =\sum_{k=0}^{N}p(t_k)q(t_k)$.

\begin{figure}[ht]
 \centering
		\begin{tabular}{cc}
     	\subfigure[$\ $Uniform Grid]{
	            \label{fig:uniform}
	            \includegraphics[width=.5\linewidth]{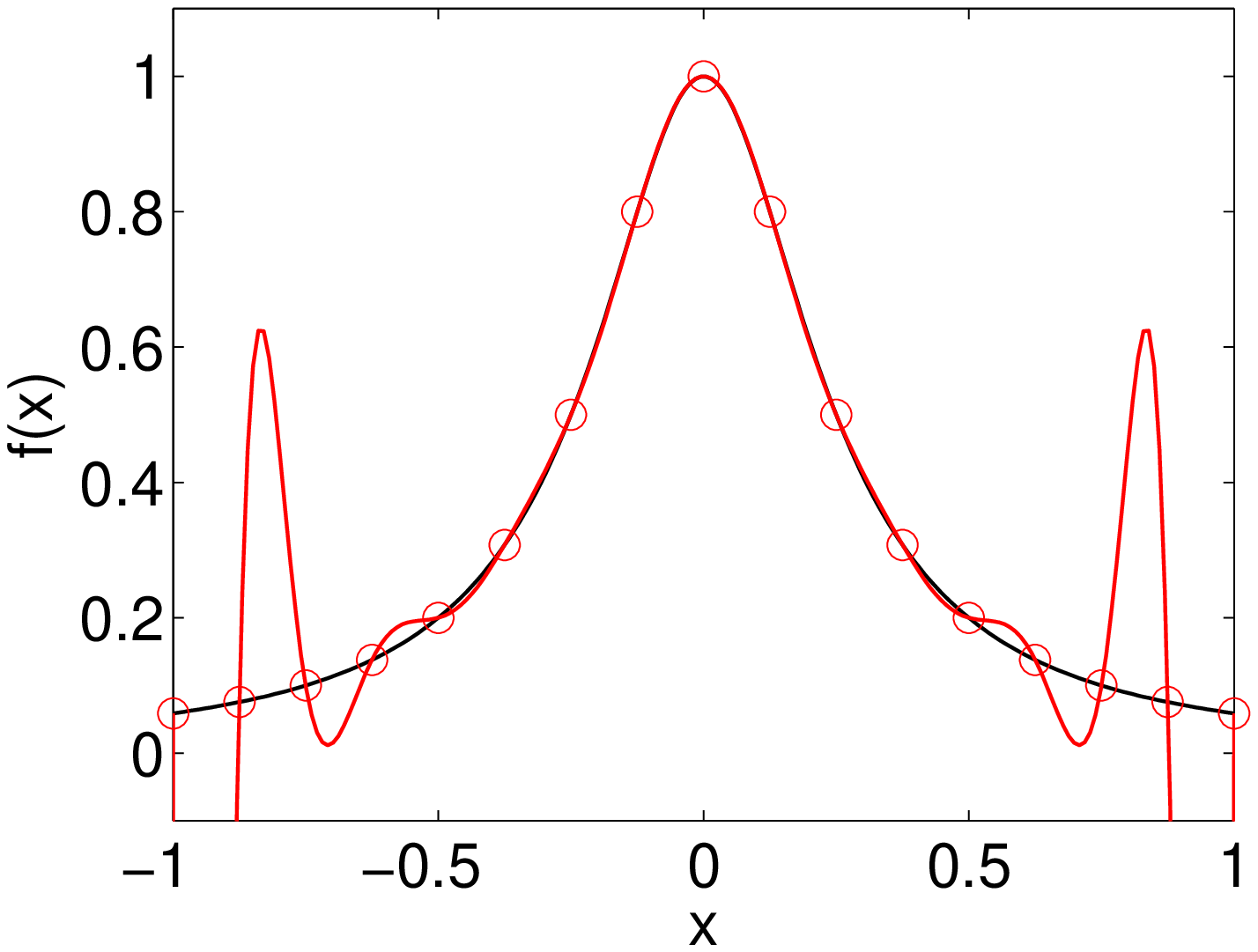}} &
	        \subfigure[$\ $LGL Grid]{
	            \label{fig:lgl}
	            \includegraphics[width=.5\linewidth]{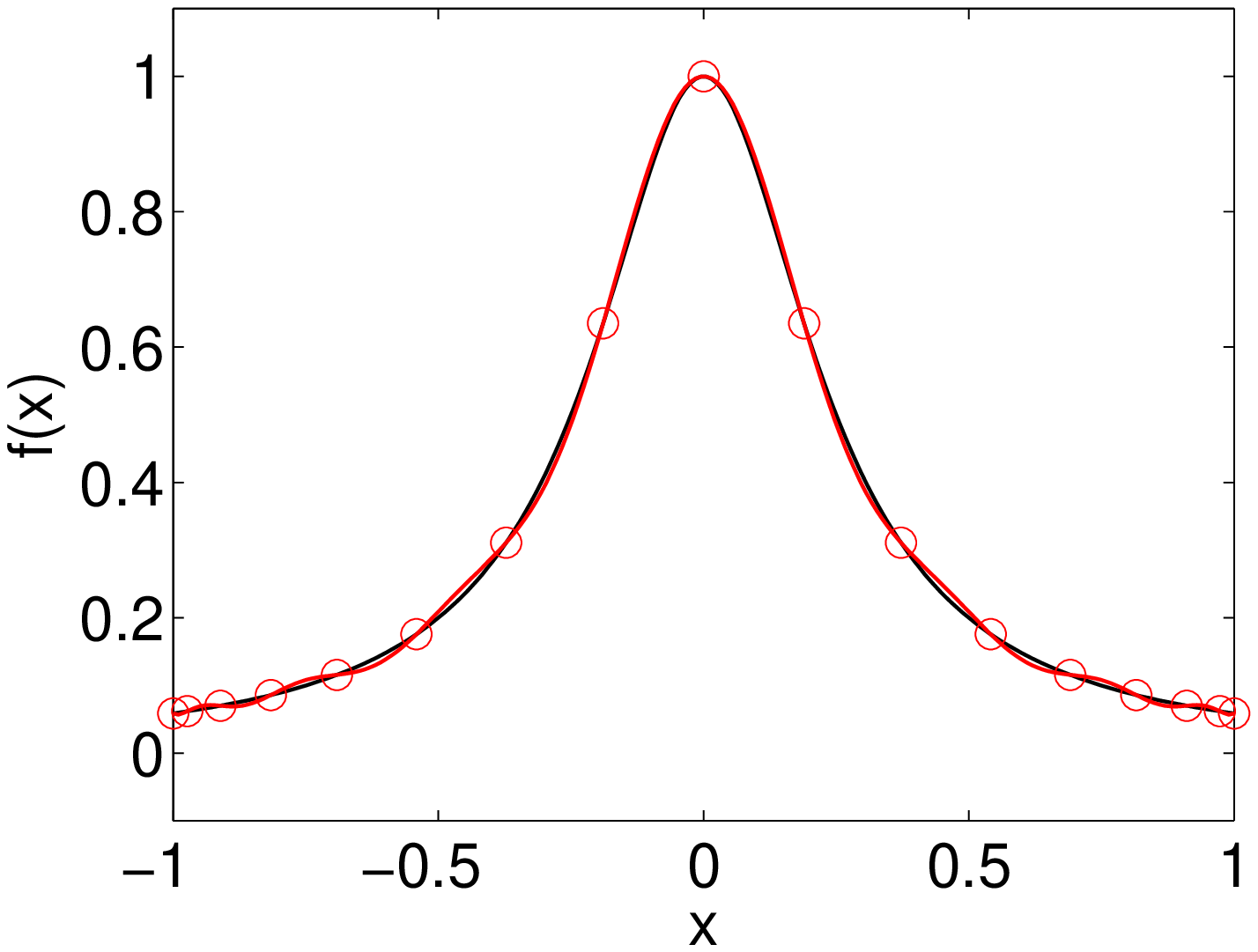}}
		\end{tabular}
 \caption{The $N=16$ order interpolation of the function $f(x)=1/(16x^2+1)$ based on a uniform grid (panel a) demonstrates the Runge phenomenon whereas the interpolation based on the LGL grid (panel b) does not.}
 \label{fig:uniform_lgl}
\end{figure}

With these tools, we can now write the $N^{\text{th}}$ order interpolating approximations of the state trajectory and control functions with respect to a given grid $\Gamma$ of $N+1$ nodes as,
\begin{eqnarray} 
	\label{eq:Ix} x(t) &\approx I_N x(t) = \sum_{k=0}^N \bar{x}_k \ell_k(t), \\
	\label{eq:Iu} u(t) &\approx I_N u(t) = \sum_{k=0}^N \bar{u}_k \ell_k(t),
\end{eqnarray}
where $\bar{x}_k$ and $\bar{u}_k$ are not only the coefficients of the expansions, but also the function values at the $k^{\text{th}}$ node due to the definition of the Lagrange polynomials \cite{elnagar_pseudospectral_1995}. Because these Lagrange polynomials are constructed based on the choice of these nodes, the approximations made with this basis as in \eqref{eq:Ix} and \eqref{eq:Iu} are sensitive to the choice of the nodes. For an arbitrary selection of nodes, as the order of approximation $N$ gets large, Runge phenomenon may occur, that is, there are increasingly larger spurious oscillations near the endpoints of the $[-1,1]$ domain \cite{fornberg_practical_1998} as shown in Fig. \ref{fig:uniform_lgl}. A selection of Gauss-type nodes with quadratic spacing towards the endpoints suppresses such oscillation between the interpolation nodes and greatly increases the accuracy of the approximation \cite{boyd_chebyshev_2000}. It has been shown that for a fixed $N>0$ and a norm given by (\ref{eq:PSinfnorm}), Gauss-type nodes are asymptotically close to optimal for interpolating a continuous function over the domain $[-1,1]$ \cite{smith_lebesgue_2006}.

In order to maintain the advantages of a spectral method while using collocation, we write the Lagrange polynomials in terms of orthogonal polynomials. We choose to focus on the Legendre polynomials which are orthogonal in $L_2[-1,1]$ defined with a weighted inner product,
$$\langle f,g \rangle = \int_{-1}^1 f(t)g(t)w(t)dt,$$
with a constant weight function $w(t)=1$, $\forall\,t\in[-1,1]$, where $f,g\in L_2[-1,1]$. Implementing the pseudospectral method with the orthogonal Legendre polynomials determines the grid to be Legendre-Gauss nodes which are the roots of $\dot{L}_N(t)$, the derivative of the $N^{\rm th}$ order Legendre polynomial. To enforce the method at the end points, we use the Legendre-Gauss-Lobotto (LGL) nodes, which include $t_0=-1$ and $t_N=1$, i.e., $\Gamma^{LGL}=\{t_j:\dot{L}_N(t)|_{t_j}=0, j=1,\ldots N-1\} \bigcup \{-1,1\}$. Then, the Lagrange polynomials $\ell_k(t)$ can be expressed with respect to $\Gamma^{LGL}$ as \cite{szego_orthogonal_1959}
\begin{equation*}\label{eq:lagleg}
	\ell_k(t) = \displaystyle\frac{1}{N(N+1)L_N(t_k)}\frac{(t^2-1) \dot{L}_N(t)}{t-t_k},
\end{equation*}
where $\{t_k\}\in\Gamma^{LGL}$, $k=0,1,\ldots,N$.

From the interpolation as in \eqref{eq:Ix}, we have
\begin{equation*}\label{eq:dinterpsc}	
\frac{d}{dt} I_N x(t) = \sum_{k=0}^N \bar{x}_k \dot{\ell}_k(t).
\end{equation*}
Using special recursive identities for the derivative of Legendre polynomials \cite{williams_gauss--lobatto_2006}, we have at the LGL nodes $t_j\in\Gamma^{LGL}$, $j=0,1,\ldots,N$,
\begin{equation}\label{eq:dinterpsc_j}
	\frac{d}{dt} I_N x(t_j) = \sum_{k=0}^N \bar{x}_k \dot{\ell}_k(t_j)=\sum_{k=0}^N D_{jk}\bar{x}_k,
\end{equation}
where $D_{jk}$ are $jk^{\rm th}$ elements of the constant $(N+1)\times(N+1)$ differentiation matrix $D$ defined by \cite{gottlieb_theory_1984}
\begin{equation}\label{eq:D}
D_{jk} = \left\{
\begin{array}{cl}
    \frac{L_N(t_j)}{L_N(t_k)}\frac{1}{t_j-t_k} & j \neq k \\  & \\
    -\frac{N(N+1)}{4} & j=k=0 \\ & \\
    \frac{N(N+1)}{4} & j=k=N \\ & \\
    0 & \textrm{otherwise}.
\end{array}
\right.
\end{equation}

In addition, the integral cost functional in the optimal control problem \eqref{eq:opc-11} can be approximated by Gaussian quadrature. In particular, Legendre-Gauss-Lobotto quadrature is used to enforce endpoint conditions and defined as
\begin{equation}\label{eq:gaussquad}
	\int_{-1}^{1} f(t) dx = \sum_{i=1}^{N} f(t_i) w_i, \qquad w_i = \int_{-1}^1 \ell_i(t)dt,
\end{equation}
which is exact for $f\in\mathbb{P}_{2N-1}$ when $\{t_i\}\in\Gamma^{LGL}$ \cite{canuto_spectral_2006}. Therefore, the choice of LGL nodes not only achieves close to optimal interpolation error by preventing increasingly spurious oscillations as $N$ gets large but also ensures the accuracy of the numerical integration.

Compiling equations (\ref{eq:Ix}), (\ref{eq:dinterpsc_j}), and (\ref{eq:gaussquad}) we can convert the optimal control problem as in \eqref{eq:opc-11} into the following finite-dimensional constrained minimization problem by discretizing the states and controls with an interpolation scheme, representing the differential equation through recursive definition of spectral derivatives, and expressing integral terms with Gaussian quadrature,
\begin{align*}
\min\ \ & \varphi(T,\bar{x}_N)+\frac{T}{2}\sum_{i=0}^N \mathcal{L}(\bar{x}_i,\bar{u}_i)w_i \\
{\rm s.t.}\ \ & \sum_{k=0}^N D_{jk} \bar{x}_k =\frac{T}{2}\Big[\mathcal{H}_f+\sum_{i=1}^{m}\bar{u}_{ij} \mathcal{H}_i\Big]x_j,\\
& e(\bar{x}_0,\bar{x}_N) = 0,\\
& g(\bar{x}_j,\bar{u}_j)\leq 0, \quad \forall\ j\in\{0,1,\ldots,N\},
\end{align*}
where $\bar{u}_{ij}, i=1,\ldots,m$, are components of the vector $\bar{u}_j$ denoting the value of the control function $u_i$ at the $j^{\rm th}$ LGL node $t_j$, namely, $\bar{u}_j=(\bar{u}_{1j},\ldots,\bar{u}_{mj})^T=(u_1(t_j),\ldots,u_m(t_j))^T$. Solvers for this type of constrained minimization problem are readily available and straight-forward to implement.

\section{Examples from Nuclear Magnetic Resonance Spectroscopy in Liquids}
\label{sec:example}
In this section we show the robustness and efficiency of the pseudospectral method by applying it to a series of optimal control problems on open quantum systems that arise in NMR spectroscopy of proteins in liquids. These control problems were selected because analytical expressions for their optimal solutions have been derived in the literature \cite{Khaneja03_1,Khaneja03_2,Stefanatos05}, making them well suited for testing the performance of the pseudospectral method on open quantum systems.

\subsection{Pair of Coupled Heteronuclear Spins}
The first open quantum system from liquid state NMR that we consider is an isolated pair of heteronuclear spins $1/2$ (spins that belong to different nuclear species), denoted as $I_1$ (for example $^1$H) and $I_2$ (for example $^{13}$C or $^{15}$N), with a scalar coupling $J$ \cite{Goldman88}. In a doubly rotating frame, which rotates with each spin at its resonance (Larmor) frequency, the free evolution Hamiltonian for this system is $H_{f}=2JI_{1z}I_{2z}$, where $I_{1z}=\sigma_{1z}/2,I_{2z}=\sigma_{2z}/2$ and $\sigma_{1z}, \sigma_{2z}$ are the Pauli spin matrices for spins $I_1$ and $I_2$ respectively. Note that this Hamiltonian is valid in the so-called weak coupling limit, where the resonance frequencies of the spins satisfy $|\omega_1-\omega_2| \gg J$ and thus the Heisenberg coupling ($\textit{\textbf{I}}_1\cdot\textit{\textbf{I}}_{2}$), which is the characteristic indirect coupling between spins in isotropic liquids, can be approximated by the scalar coupling ($I_{1z}I_{2z}$) \cite{Goldman88}.

The most important relaxation mechanisms in NMR spectroscopy in liquid solutions are due to Dipole-Dipole (DD) interaction and Chemical Shift Anisotropy (CSA), as well as their interference effects, i.e., DD-CSA cross correlation \cite{Goldman88}. We initially consider the spin system without cross-correlated relaxation.

\subsubsection{Spin Pair without Cross-Correlated Relaxation}
\label{sec:nocross}
Here we consider the open quantum system with only DD and CSA relaxation ignoring the cross-correlated relaxation. This case approximates for example the situation for deuterated and $^{15}$N-labeled proteins in ${\rm H_2O}$ at moderately high magnetic fields (e.g., 10 Tesla), where the $^1$H-$^{15}$N spin pairs are isolated and CSA relaxation is small. Furthermore, we focus on slowly tumbling molecules in the so-called spin diffusion limit \cite{Ernst87}. In this case, longitudinal relaxation rates ($1/T_1$) are negligible compared to transverse relaxation rates ($1/T_2$) \cite{Ernst87}.

For this coupled two-spin system, the free evolution of the density matrix $\rho$ in the doubly rotating frame is given by the following master equation \cite{Khaneja03_1}
\begin{align}
\label{ro1}
\dot{\rho}= &-iJ[2I_{1z}I_{2z},\rho]-k_{DD}[2I_{1z}I_{2z}, [2I_{1z}I_{2z}, \rho\,]]\nonumber\\
& -k^{1}_{CSA}[I_{1z}, [I_{1z}, \rho\,]]-k^{2}_{CSA}[I_{2z}, [I_{2z}, \rho\,]],
\end{align}
where $J$ is the scalar coupling constant, $k_{DD}$ is the DD relaxation rate, and $k^{1}_{CSA}, k^{2}_{CSA}$ are CSA relaxation rates for spins $I_1, I_2$, respectively . These relaxation rates depend on various physical parameters of the system, like the gyromagnetic ratios of the spins, the internuclear distance and the correlation time of the rotational tumbling \cite{Goldman88}.

One problem of interest in NMR experiments is to find the pulses (controls), $\omega_x(t)$ and $\omega_y(t)$ applied in the $x$ and $y$ directions, respectively, for optimal polarization transfer $I_{1z}\rightarrow I_{2z}$ from one spin to the other. This transfer is suitably done in two steps: $I_{1z}\rightarrow 2I_{1z}I_{2z}\rightarrow I_{2z}$. Since these two steps are symmetric, the optimal controls for the second step are symmetric to those of the first one. Thus, we only need to concentrate on the first step and the objective is to maximize the transfer $I_{1z}\rightarrow 2I_{1z}I_{2z}$. In other words, starting from the initial state $\rho(0)=I_{1z}$, we tend to maximize the final expectation value of the target operator $O=2I_{1z}I_{2z}$, i.e., $\langle 2I_{1z}I_{2z}\rangle(T)=\mbox{trace} \{ \rho(T) 2I_{1z}I_{2z}\}$, where $T$ is the final time of the experiment and $\rho(T)$ is controlled by $\omega_x(t)$ and $\omega_y(t)$. Using the master equation (\ref{ro1}), we can find differential equations that describe the time evolution of the expectations of the operators participating in the desired transfer as presented in Section \ref{sec:ps}. The corresponding equations in matrix form are
\begin{equation}
\label{path1}
\left[\begin{array}{cccc}
\dot{x}_1\\\dot{x}_2\\\dot{x}_3\\\dot{x}_4\end{array}\right]
=\left[\begin{array}{cccc} 0 & -u_1 & 0 & 0 \\
u_1 & -\xi & -1 & 0 \\
0 & 1 & -\xi & -u_2 \\
0 & 0 & u_2 & 0 \\
\end{array}\right]
\left[\begin{array}{cccc}
x_1\\x_2\\x_3\\x_4\end{array}\right],
\end{equation}
where $x_1=\langle I_{1z}\rangle$, $x_2=\langle I_{1x}\rangle$, $x_3=\langle 2I_{1y}I_{2z}\rangle$, $x_4=\langle 2I_{1z}I_{2z}\rangle$, $\xi=(k_{DD}+k^{1}_{CSA})/J$, and the controls $u_1(t)=\omega_y(t)/J$ and $u_2(t)=\omega_x(t)/J$ are the normalized (w.r.t. $J$) transverse components of the applied magnetic field. Note that the above system (\ref{path1}) has the bilinear form as shown in (\ref{bilinear}).

Consequently, we now arrive at an optimal control problem for the transfer $I_{1z}\rightarrow 2I_{1z}I_{2z}$ that is to find $u_1(t)$ and $u_2(t)$, $0\leq t \leq T$, such that starting from $x(0)=(1, 0, 0, 0)^T$, $x_4(T)$ is maximized subject to the evolution equations \eqref{path1}. This problem has been solved analytically and the resulting analytical pulse was denoted as ROPE \cite{Khaneja03_1}. It is shown there that the maximum achievable value of $x_4$, i.e., the efficiency $\eta_1$ of the transfer is given as a function of parameter $\xi$ by
\begin{equation}
\label{eta1} \eta_1= \sqrt{\xi^2+1}-\xi\;.
\end{equation}

\begin{figure}[ht]
\centering
\includegraphics[scale=0.5]{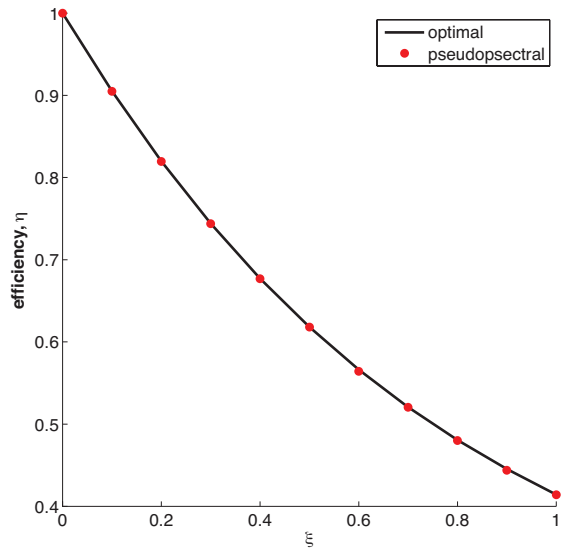}
\caption{The efficiency of the transfer $x_1\rightarrow x_4$ in system (\ref{path1}) achieved by the pseudospectral method, as a function of the relaxation parameter $\xi$ in the range $[0,1]$. The theoretically calculated maximum efficiency given by (\ref{eta1}) is also shown.}
\label{fig:rope}
\end{figure}

Using the pseudospectral method presented in this paper, we calculated numerically optimal controls $u_1(t), u_2(t)$ for various values of $\xi$ in the range $[0,1]$ to maximize the corresponding achievable values of $x_4(T)$. The method was implemented in Matlab using the third party KNITRO nonlinear programming solver from Ziena Optimization. The problem was approximated using 25 ($N=24$) nodes and with the terminal time free to vary, but with a maximum time of $T=10$. A unified and general method should not use any prior knowledge, so the solver was given an arbitrary initial guess for the controls. In this case, we take $u_1(t)=u_2(t)=1$ and $T=1$. The termination tolerance on the cost function of the solver was set at $1\times10^{-8}$. In Fig. \ref{fig:rope} we plot the value $x_4(T)$ achieved by the pseudospectral method for $\xi \in [0,1]$. For comparison, we also plot the maximum efficiency given in (\ref{eta1}). The excellent agreement shows the efficiency of the method to approximate optimal solutions.

\begin{figure}[ht!]
 \centering
		\begin{tabular}{cc}
     	\subfigure[$\ $ROPE Control Amplitude]{
	            \label{fig:rope_controls}
	            \includegraphics[width=.45\linewidth]{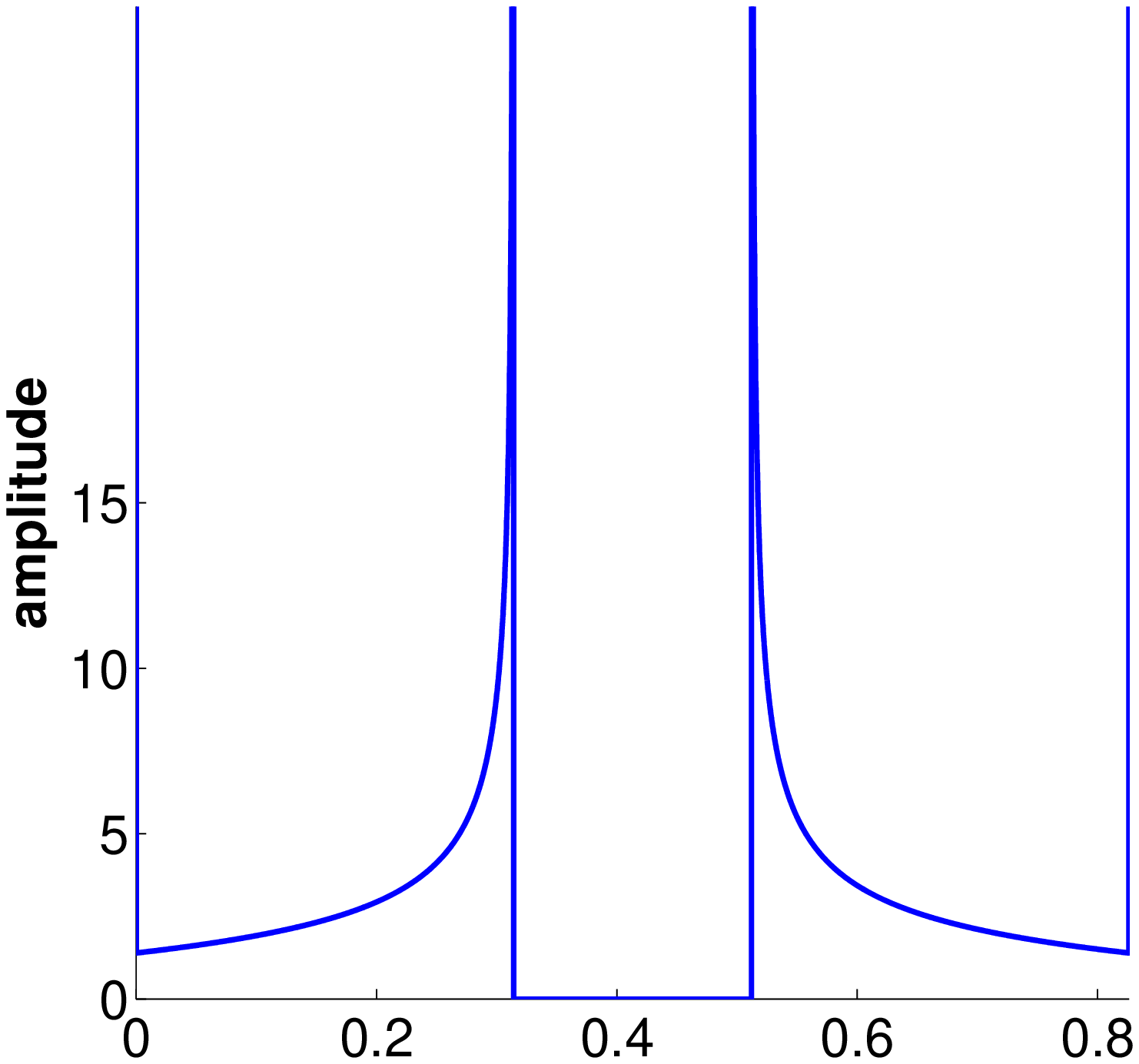}} &
	        \subfigure[$\ $PS Control Amplitude]{
	            \label{fig:ps_controls}
	            \includegraphics[width=.45\linewidth]{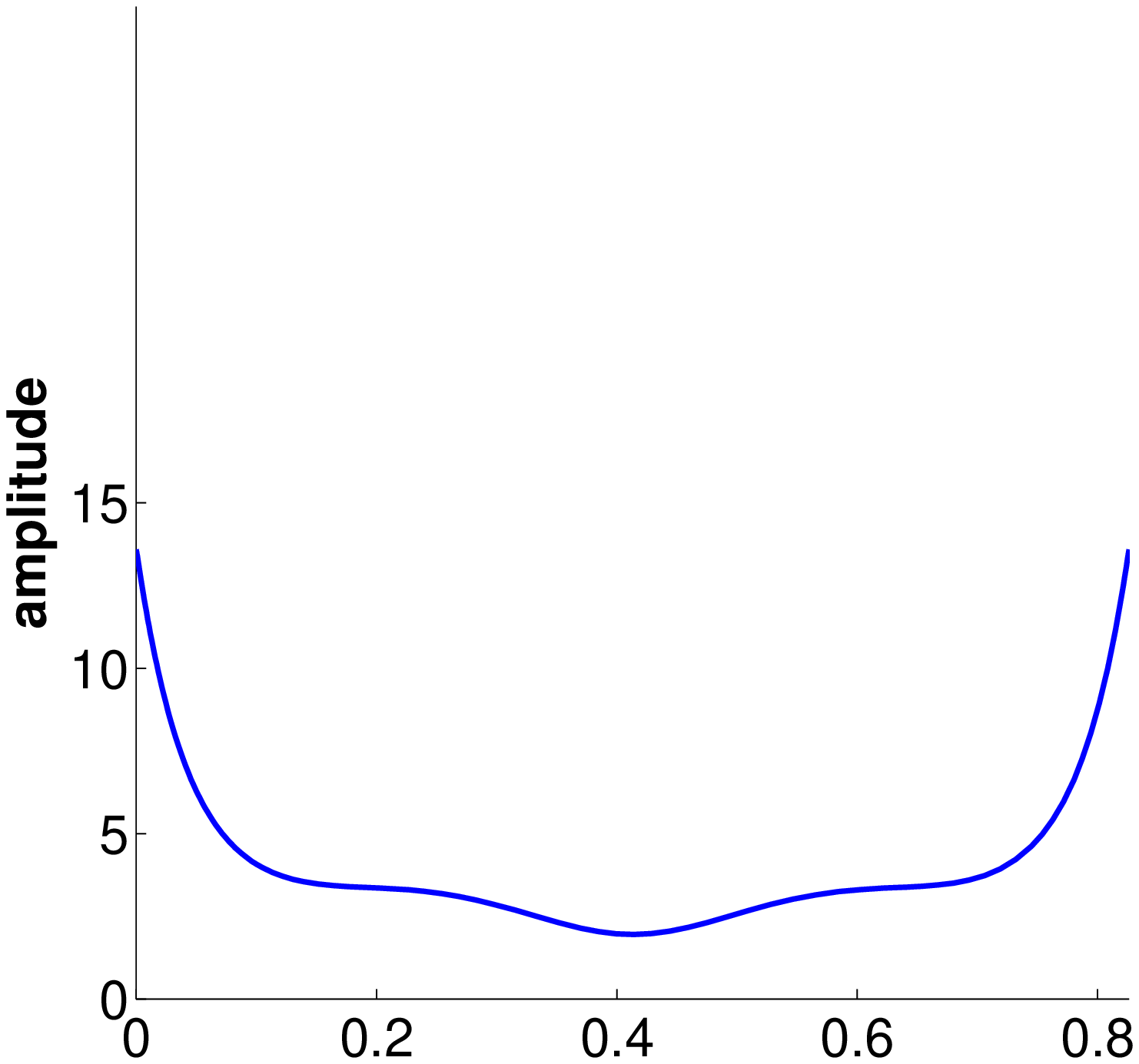}} \\
	        \subfigure[$\ $ROPE Trajectories]{
	            \label{fig:rope_traj}
	            \includegraphics[width=.45\linewidth]{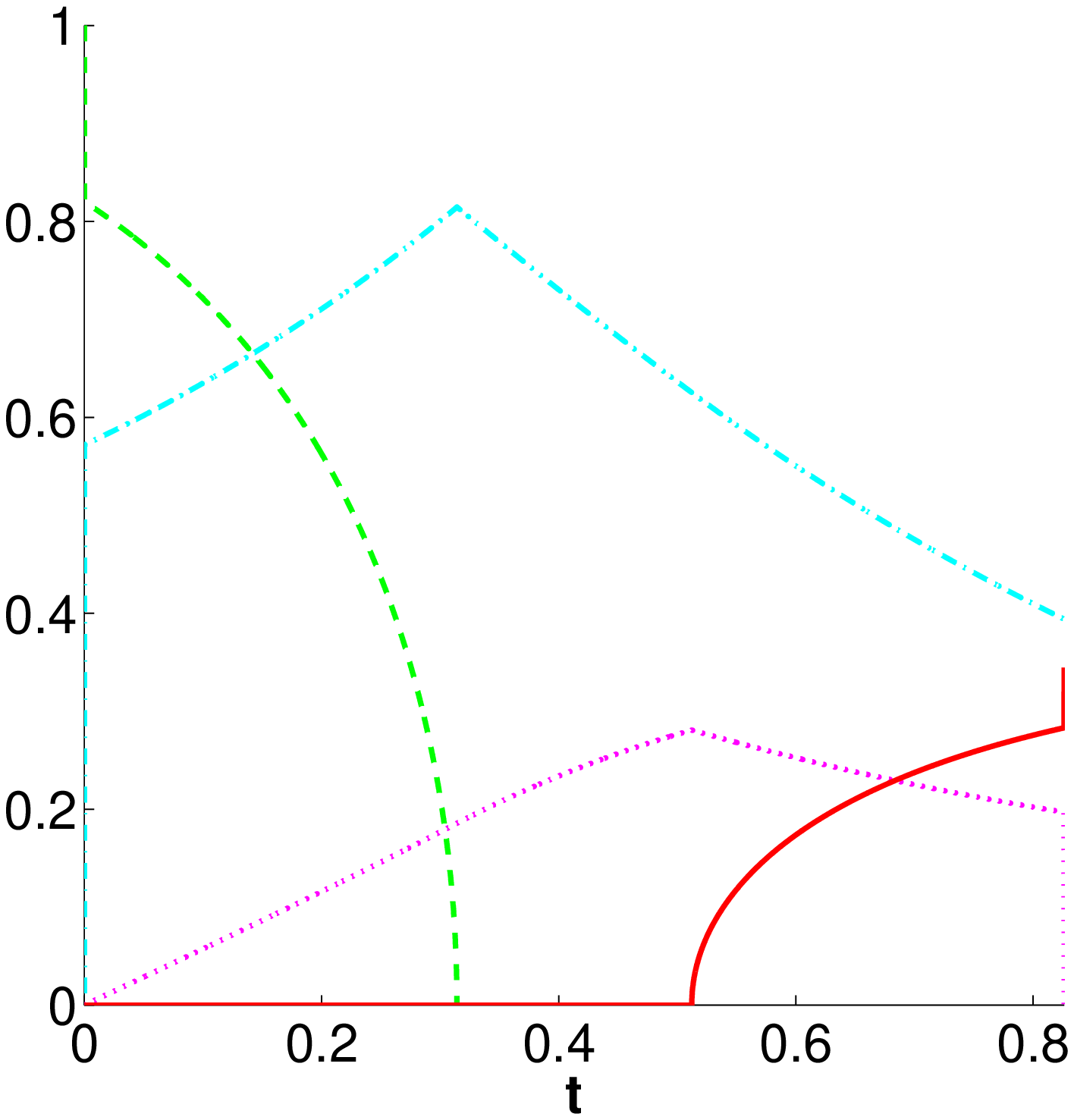}} &
			\subfigure[$\ $PS Trajectories]{
	            \label{fig:ps_traj}
	            \includegraphics[width=.45\linewidth]{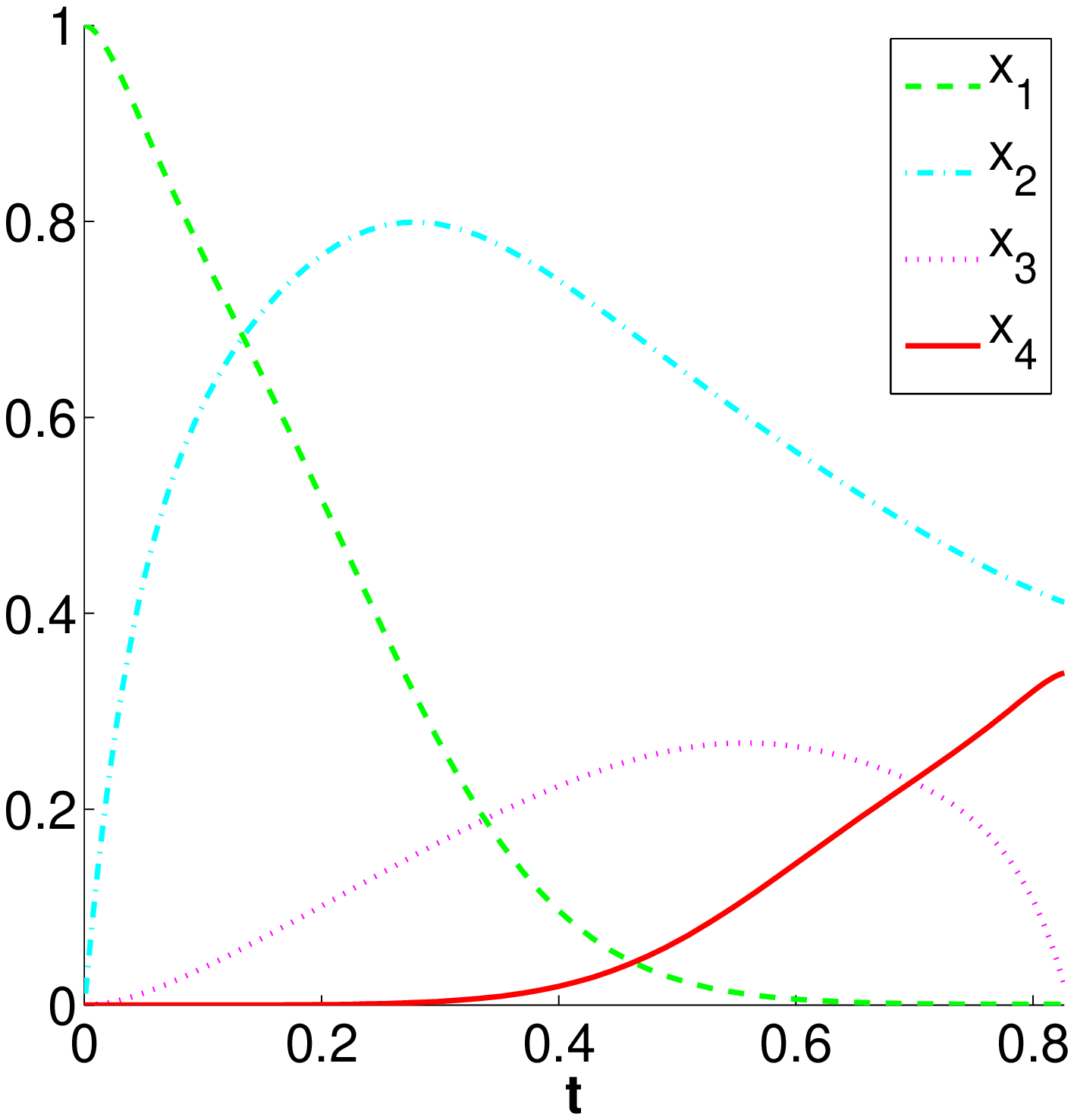}} \\
		\end{tabular}
 \caption{Pseudospectral controls (panel b) and state trajectories (panel d) are compared to analytic ROPE \cite{Khaneja03_1} controls (panel a) and trajectories (panel c) for $\xi=1$.}
 \label{fig:rope_ps}
\end{figure}

Another clear advantage of the pseudospectral method well illustrated by this problem is that the derived control pulses are smooth functions. Fig. \ref{fig:rope_controls} shows the discontinuities of the theoretically calculated optimal pulse amplitude ($\sqrt{u_1^2(t)+u_2^2(t)}$) \cite{Khaneja03_1}. Such discontinuities can be challenging, if not impossible, to implement in practice and high amplitudes can be hazardous for the experiment sample, equipment, and human subjects as in Magnetic Resonance Imaging (MRI). The pulse amplitude derived by the pseudospectral method, shown in Fig. \ref{fig:ps_controls}, is easily implementable and maintains low values despite achieving transfer efficiencies within $1\times10^{-3}$ of the theoretical optimal values. The pseudospectral pulse shown in Fig. \ref{fig:ps_controls} is attained from an optimization that minimizes energy while maintaining a desired efficiency transfer (i.e., the value of $x_4(T)$ found without taking energy considerations into account). Since there are many pulses that achieve this near-optimal transfer this optimization selects the pulse with minimum energy. Therefore, not only is the pseudospectral pulse without discontinuities but it also accomplishes the transfer with 45\% less energy than the ROPE pulse.

\subsubsection{Spin Pair with Cross-Correlated Relaxation}
\label{sec:cross}
If DD-CSA cross-correlated relaxation cannot be neglected, the master equation as in (\ref{ro1}) is then modified to incorporate it as \cite{Khaneja03_2}
\begin{align*}
\label{ro2}
\dot{\rho}=&-iJ[2I_{1z}I_{2z},\rho]-k_{DD}[2I_{1z}I_{2z}, [2I_{1z}I_{2z}, \rho\,]]\\
&-k^{1}_{CSA}[I_{1z}, [I_{1z}, \rho\,]]-k^{2}_{CSA}[I_{2z}, [I_{2z}, \rho\,]]\\
&-k^{1}_{DD/CSA}[2I_{1z}I_{2z}, [I_{1z}, \rho\,]]\\
&-k^{2}_{DD/CSA}[2I_{1z}I_{2z}, [I_{2z}, \rho\,]],
\end{align*}
where $k_{DD}, k^{1}_{CSA}, k^{2}_{CSA}$ are auto-relaxation rates due to DD relaxation, CSA relaxation of spin $I_1$, CSA relaxation of spin $I_2$ and $ k^{1}_{DD/CSA}, k^{2}_{DD/CSA}$ are cross-correlation rates of spins $I_1$ and $I_2$ due to interference effects between DD and CSA relaxation mechanisms.

Using this master equation, we can find the following equations for the ensemble averages
\begin{equation}
\label{path2} \left[\begin{array}{cccccc}
\dot{x}_1\\\dot{x}_2\\\dot{x}_3\\\dot{x}_4\\\dot{x}_5\\\dot{x}_6\end{array}\right]
=\left[\begin{array}{cccccc} 0 & -u_1 & u_2 & 0 & 0 & 0 \\
u_1 & -\xi_a & 0 & -1 & -\xi_c & 0 \\
-u_2 & 0 & -\xi_a & -\xi_c & 1 & 0 \\
0 & 1 & -\xi_c & -\xi_a & 0 & -u_2 \\
0 & -\xi_c & -1 & 0 & -\xi_a & u_1 \\
0 & 0 & 0 & u_2 & -u_1 & 0 \\
\end{array}\right]
\left[\begin{array}{cccccc}
x_1\\x_2\\x_3\\x_4\\x_5\\x_6\end{array}\right],
\end{equation}
where $x_1=\langle I_{1z}\rangle$, $x_2=\langle I_{1x}\rangle$, $x_3=\langle I_{1y}\rangle$, $x_4=\langle 2I_{1y}I_{2z}\rangle$, $x_5=\langle 2I_{1x}I_{2z}\rangle$, $x_6=\langle 2I_{1z}I_{2z}\rangle$, $\xi_a=(k_{DD}+k^{1}_{CSA})/J$, $\xi_c=k^{1}_{DD/CSA}/J$ and $u_1(t), u_2(t)$ are the available controls as before. Starting from $x(0)=(1, 0, 0, 0, 0, 0)^T$, we want to design $u_1(t)$ and $u_2(t)$ that maximize $x_6(T)$ subject to \eqref{path2}.

This problem has also been solved analytically and the analytical pulse was denoted as CROP \cite{Khaneja03_2}. It is shown there that the maximum achievable value of $x_6$, i.e., the efficiency $\eta_2$ of the transfer is given by the same formula as before
\begin{equation}
\label{eta2} \eta_2= \sqrt{\xi^2+1}-\xi\;,
\end{equation}
but now
\begin{equation}
\label{zeta} \xi= \sqrt{\frac{\xi_a^2-\xi_c^2}{1+\xi_c^2}}\;.
\end{equation}

\begin{figure}[t]
\centering
\includegraphics[scale=0.5]{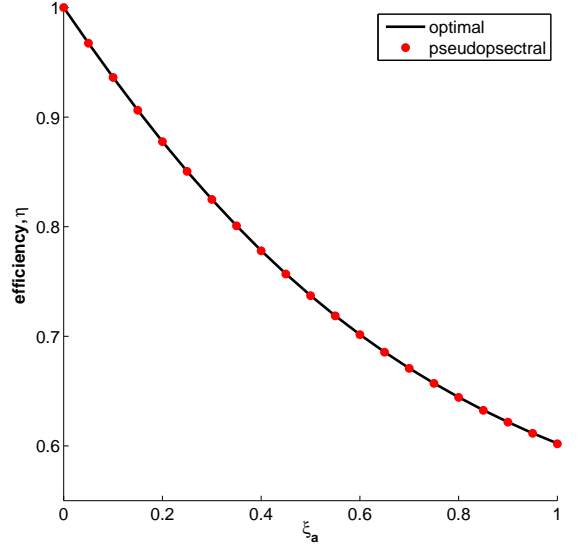}
\caption{The efficiency of the transfer $x_1\rightarrow x_6$ in system (\ref{path2}) achieved by the pseudospectral method, as a function of the relaxation parameter $\xi_a$ in the range $[0,1]$, with $\xi_c=0.75\xi_a$. The theoretically calculated maximum efficiency given by (\ref{eta2}) is also shown.}
\label{fig:crop}
\end{figure}

\begin{figure}[t]
 \centering
		\begin{tabular}{cc}
     	\subfigure[$\ $CROP Control Amplitude]{
	            \label{fig:crop_controls}
	            \includegraphics[width=.45\linewidth]{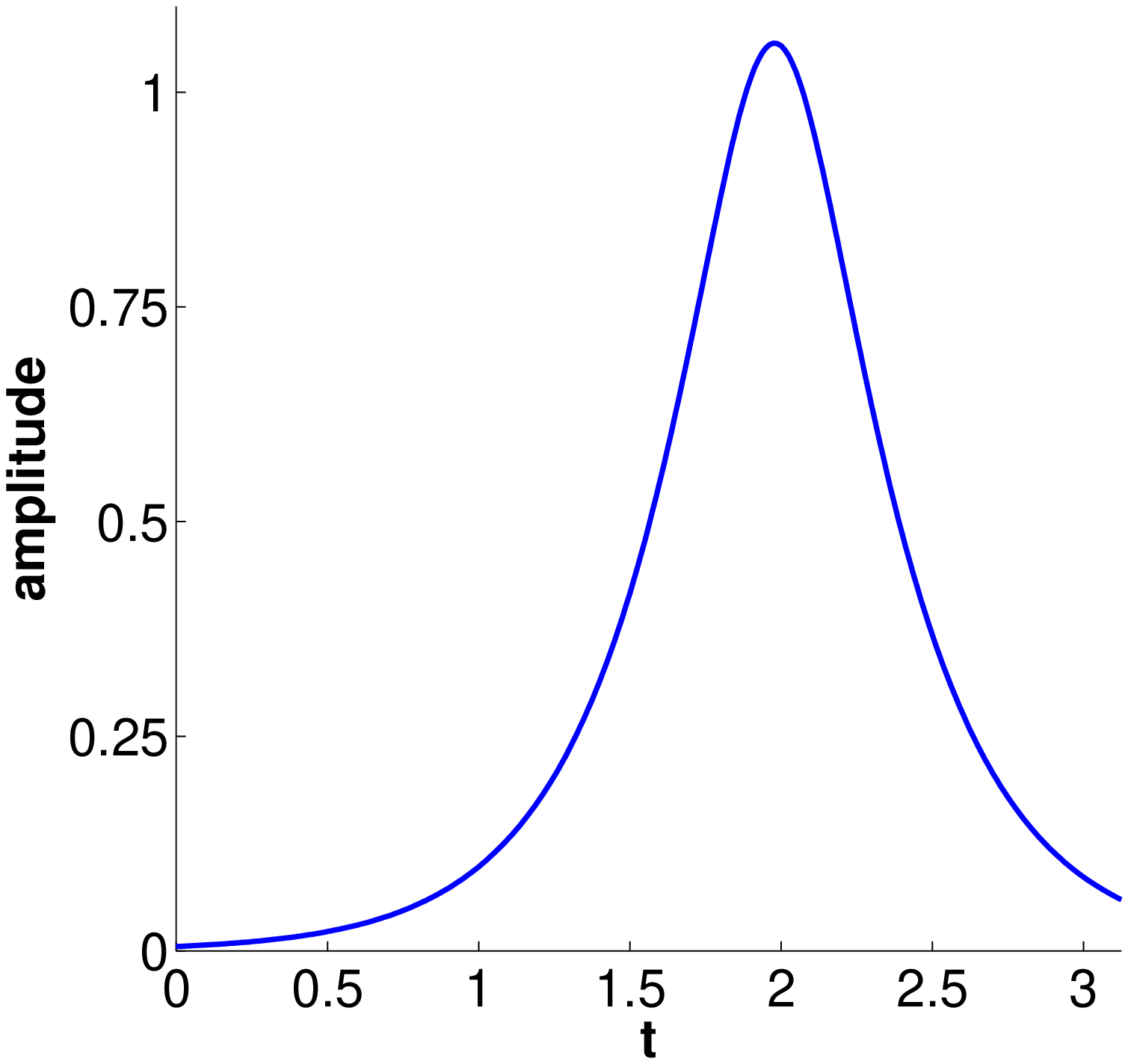}} &
	        \subfigure[$\ $PS Control Amplitude]{
	            \label{fig:ps2_controls}
	            \includegraphics[width=.45\linewidth]{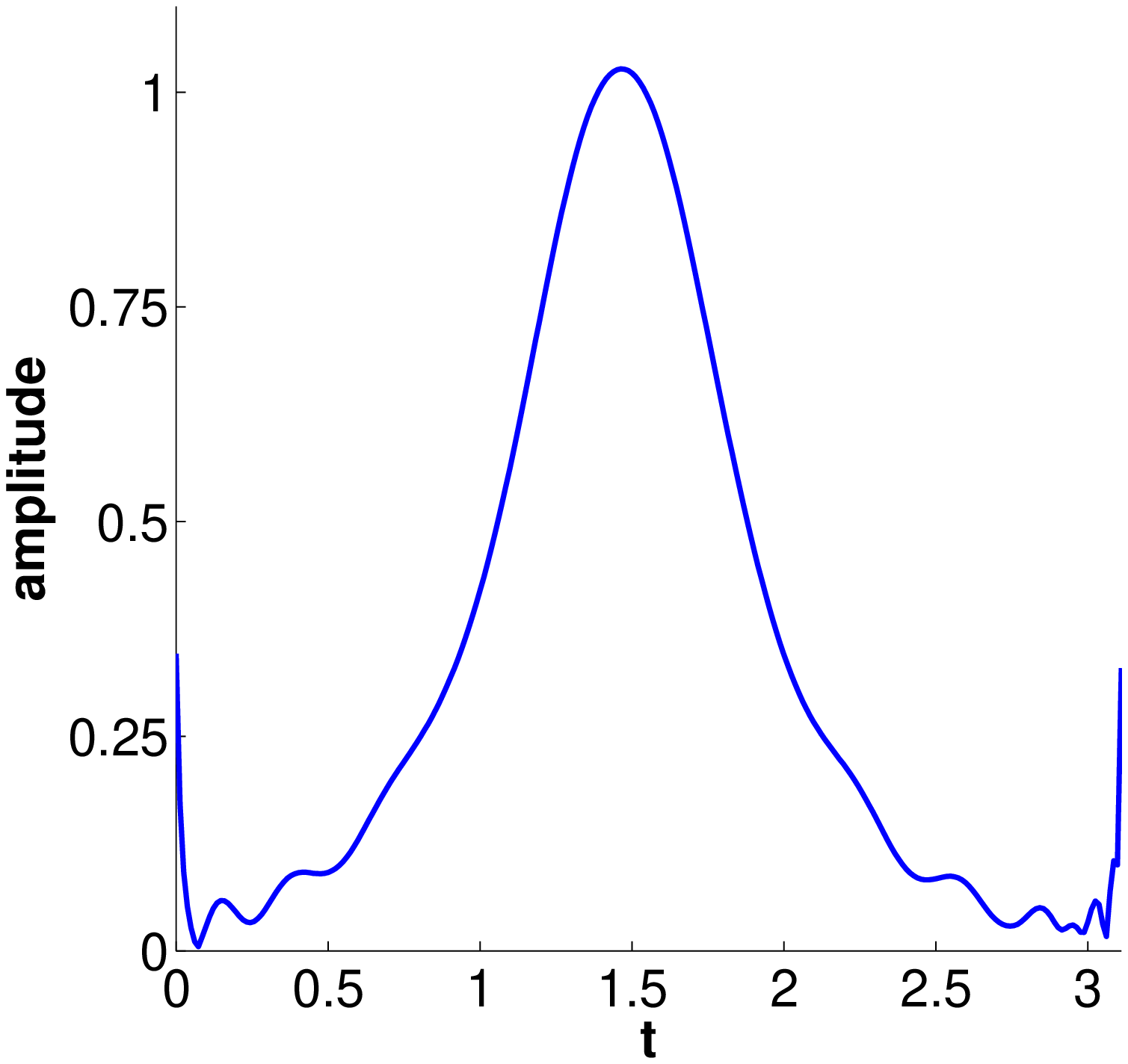}}
		\end{tabular}
 \caption{The CROP derived control pulse (panel a) is compared to the pseudospectral control pulse (panel b) for $\xi_a=1$ and $\xi_c=0.75$.}
 \label{fig:crop_ps}
\end{figure}

Using the pseudospectral method introduced in this paper, we calculated numerically optimal controls $u_1(t), u_2(t)$ for various values of $\xi_a$ over $[0,1]$ and with $\xi_c=0.75\xi_a$, to maximize the corresponding achievable values of $x_6(T)$. Using the same Matlab program and KNITRO solver, the optimal control problem was approximated by 25 ($N=24$) nodes with a free terminal time (maximum $T=5$). A similar constant initial guess was used and the cost function tolerance was set to $1\times10^{-5}$. In Fig. \ref{fig:crop} we plot the values of $x_6(T)$ achieved by the pseudospectral method and the maximum efficiency given in (\ref{eta2}). Again, an excellent agreement is observed. The CROP and pseudospectral control pulse amplitudes are plotted in Fig. \ref{fig:crop_ps}. 

\subsection{Three Spin Chain}
The next open quantum system that we consider is a three spin chain (spins $I_1, I_2, I_3$) with equal scalar couplings between nearest neighbors. In a suitably chosen (multiple) rotating frame, which rotates with each spin at its resonance (Larmor) frequency, the Hamiltonian $H_f$ that governs the free evolution of the system is $H_{f}=\sqrt{2}J(I_{1z}I_{2z}+I_{2z}I_{3z})$. The common coupling constant is written in the form $\sqrt{2}J$ for normalization reasons. As in the first example described in Section \ref{sec:nocross}, we neglect cross-correlated relaxation and focus on slowly tumbling molecules in the spin diffusion limit. The corresponding master equation is
\begin{align}
\label{ro3}
\dot{\rho}=&-i\sqrt{2}J[I_{1z}I_{2z}+I_{2z}I_{3z},\rho]-k_{DD}[2I_{1z}I_{2z}, [2I_{1z}I_{2z}, \rho\,]]\nonumber\\
&-k_{DD}[2I_{2z}I_{3z}, [2I_{2z}I_{3z}, \rho\,]]-k_{DD}[2I_{3z}I_{1z}, [2I_{3z}I_{1z}, \rho\,]]\nonumber\\
&-k^{1}_{CSA}[I_{1z}, [I_{1z}, \rho\,]]-k^{2}_{CSA}[I_{2z}, [I_{2z}, \rho\,]]\nonumber\\
&-k^{3}_{CSA}[I_{3z}, [I_{3z}, \rho\,]].
\end{align}
For this system we examine the polarization transfer from spin $I_1$ to spin $I_3$, $I_{1z}\rightarrow I_{3z}$. This transfer is suitably done in three steps, $I_{1z}\rightarrow 2I_{1z}I_{2z}\rightarrow 2I_{2z}I_{3z}\rightarrow I_{3z}$. The first and last steps are similar to the previously examined spin transfer, thus we concentrate on the intermediate step $2I_{1z}I_{2z}\rightarrow 2I_{2z}I_{3z}$, which corresponds to $\rho(0)=2I_{1z}I_{2z}$ and $O=2I_{2z}I_{3z}$. Similarly, from the master equation (\ref{ro3}), we can derive the associated differential equations which describe the time evolution of the expectation values of operators participating in this transfer,
\begin{equation}
\label{path3} \left[\begin{array}{ccccc}
\dot{x}_1\\\dot{x}_2\\\dot{x}_3\\\dot{x}_4\\\dot{x}_5\end{array}\right]
=\left[\begin{array}{ccccc} 0 & -u & 0 & 0 & 0 \\
u & -\xi & -1 & 0 & 0 \\
0 & 1 & -\xi & -1 & 0 \\
0 & 0 & 1 & -\xi & -u \\
0 & 0 & 0 & u & 0 \\
\end{array}\right]
\left[\begin{array}{ccccc}
x_1\\x_2\\x_3\\x_4\\x_5\end{array}\right]
\end{equation}
where $x_1=\langle 2I_{1z}I_{2z}\rangle$, $x_2=\langle 2I_{1z}I_{2x}\rangle$, $x_3=\langle\sqrt{2}(2I_{1z}I_{2y}I_{3z}+I_{2y}/2)\rangle$, $x_4=-\langle 2I_{2x}I_{3z}\rangle$, $x_5=\langle 2I_{2z}I_{3z}\rangle$, $\xi=(2k_{DD}+k^{2}_{CSA})/J$ \cite{Stefanatos05}. Transverse relaxation rate is normalized with respect to the scalar coupling $J$ and the normalized relaxation parameter is $\xi$. The control  function $u(t)=\omega_y(t)/J$ is the $y$-component of the applied magnetic field. The $x$-component creates an equivalent path from $x_1$ to $x_5$ and need not be considered \cite{Stefanatos05}.

The corresponding optimal control problem is to find, starting from $x(0)=(1, 0, 0, 0, 0)^T$, $u(t)$ that maximizes $x_5(T)$ subject to \eqref{path3}. It has been shown in \cite{Stefanatos05} that a strict upper bound, $\eta_3$, of the maximum achievable value of $x_5$ is characterized by $\xi$,
\begin{equation}
\label{eta3} \eta_3= \frac{(\sqrt{\xi^2+2}-\xi)^2}{2}\;.
\end{equation}
This bound was used in \cite{Stefanatos05} to evaluate the performance of a state-of-the-art gradient ascent algorithm for calculating optimal $u(t)$.

\begin{figure}[t]
\centering
\includegraphics[scale=0.5]{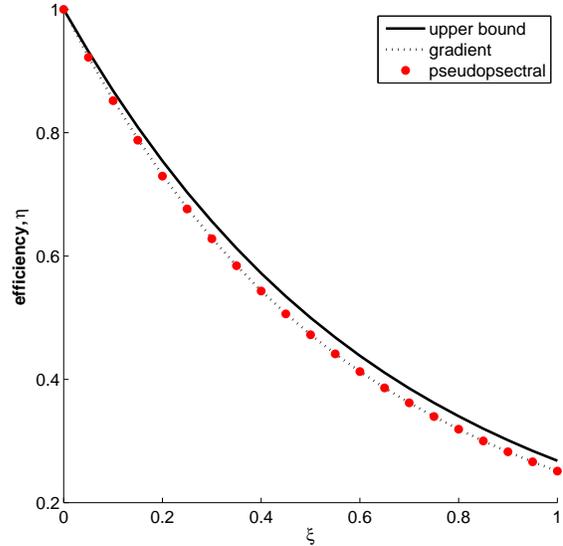}
\caption{The efficiency for transfer $x_1\rightarrow x_5$ in system (\ref{path3}) achieved by the pseudospectral method, as a function of the relaxation parameter $\xi$ in the range $[0,1]$. The corresponding numerical results achieved by a state-of-the-art gradient ascent algorithm as well as the theoretically calculated upper bound for the maximum efficiency, given by (\ref{eta3}), are also shown.}
\label{fig:chain}
\end{figure}

\begin{figure}[t]
 \centering
		\begin{tabular}{cc}
     	\subfigure[$\ $Gradient Control Amplitude]{
	            \label{fig:gradient_controls}
	            \includegraphics[width=.45\linewidth]{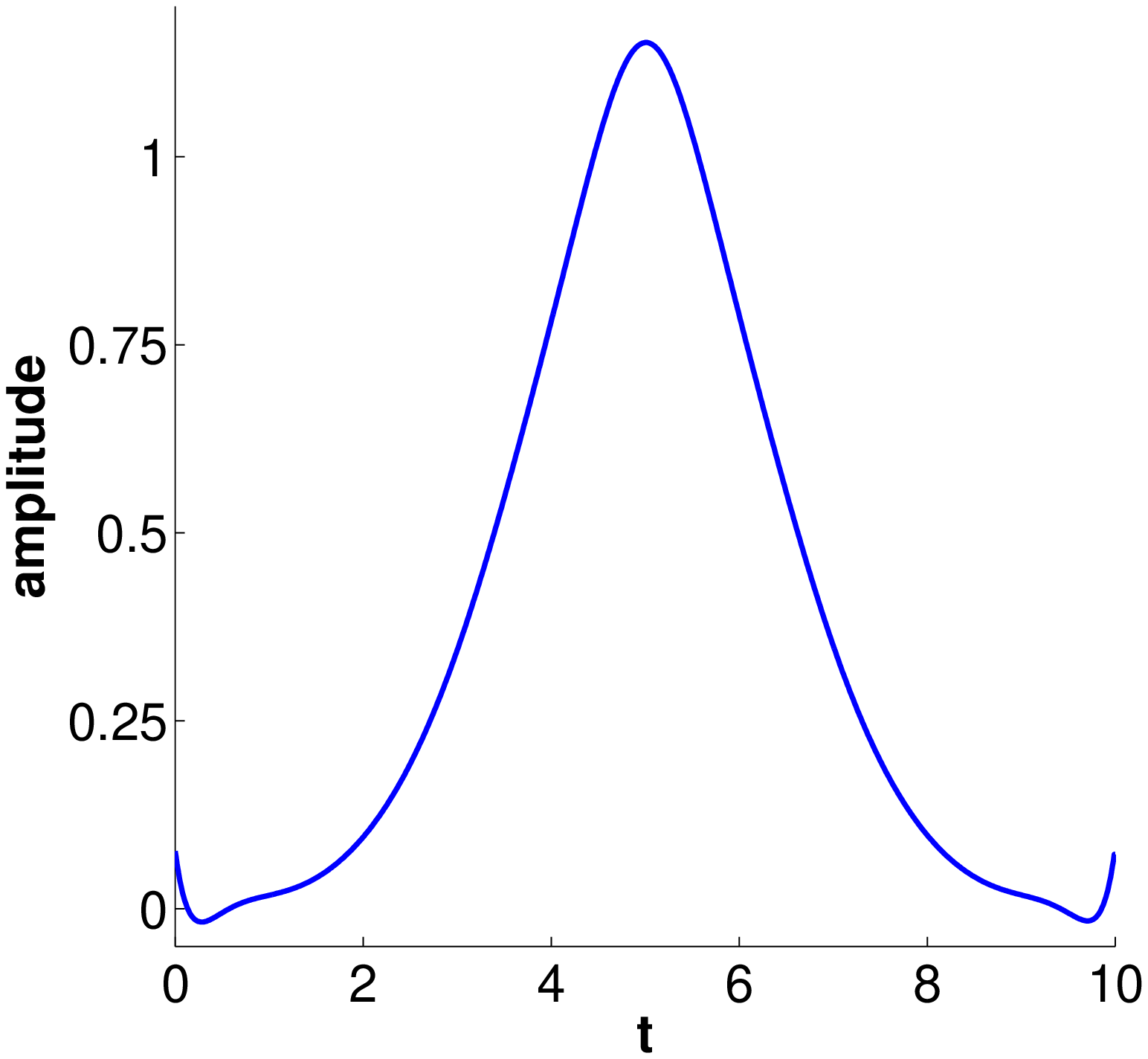}} &
	        \subfigure[$\ $PS Control Amplitude]{
	            \label{fig:ps3_controls}
	            \includegraphics[width=.45\linewidth]{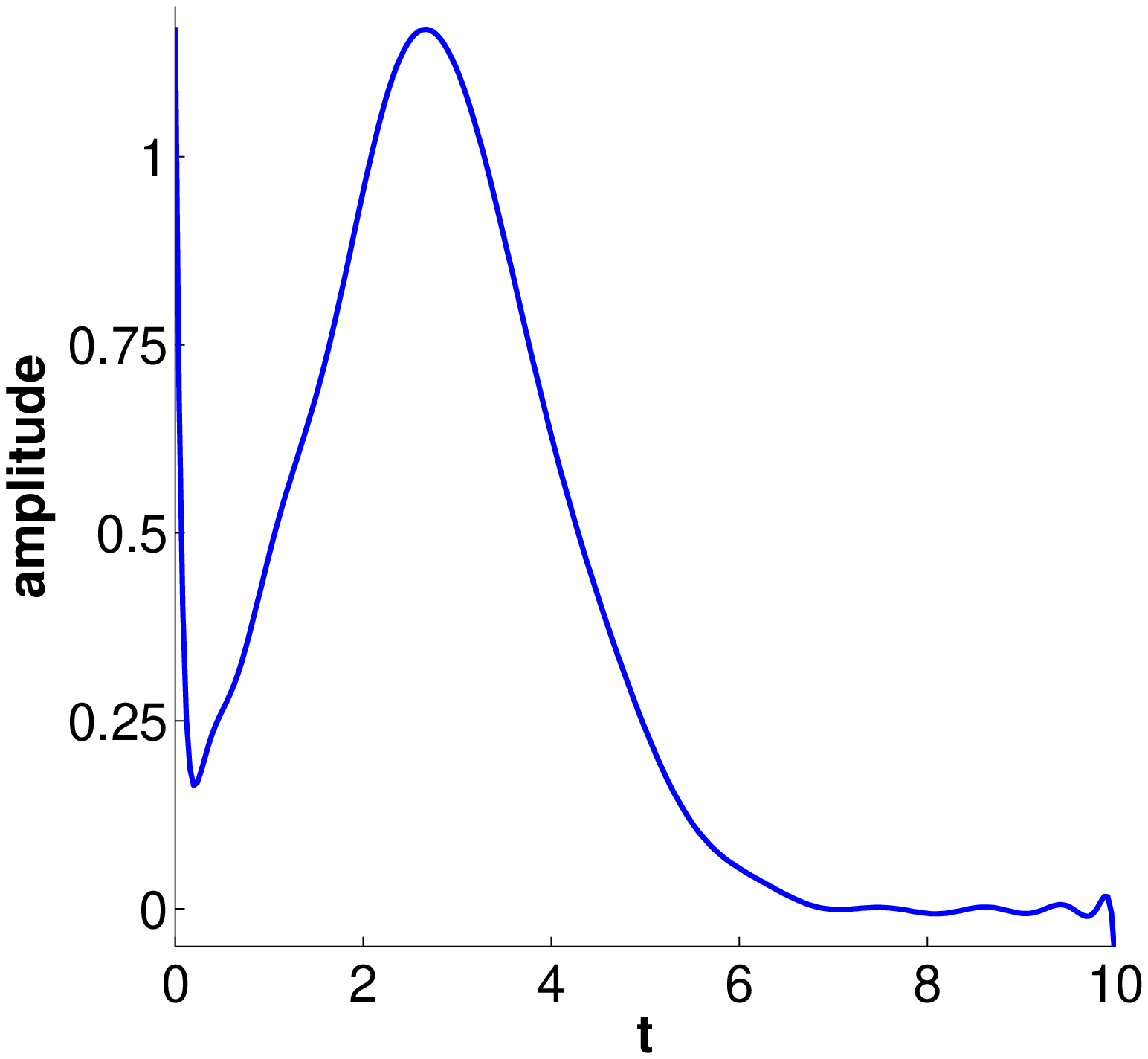}}
		\end{tabular}
 \caption{The gradient derived control pulse (panel a) is compared to the pseudospectral control pulse (panel b) for $\xi=1$.}
 \label{fig:gradient_ps}
\end{figure}

In this paper, the optimal control $u(t)$ was calculated by the pseudospectral method for various values of $\xi$ in the range $[0,1]$ to maximize the corresponding values $x_5(T)$, where the optimal control problem was approximated by 25 ($N=24$) nodes with a free terminal time (maximum $T=10$). A similar constant initial guess was used and the cost function tolerance was set to $1\times10^{-6}$. The values of $x_5(T)$ achieved by the pseudospectral method are displayed in Fig. \ref{fig:chain}. The corresponding numerical results by the state-of-the-art gradient ascent algorithm used in \cite{Stefanatos05} and the analytical efficiency upper bound given in (\ref{eta3}) are also shown in the same figure. Observe the excellent agreement between the efficiencies of the two numerical methods, lower of course than the analytical upper bound. In Fig. \ref{fig:gradient_ps} we compare the gradient control pulse with the pseudospectral control pulse for the case $\xi=1$.

\section{Conclusion}
In this paper, we presented a method based on pseudospectral approximations to effectively discretize and solve optimal control problems associated with pulse sequence design for open quantum mechanical systems. Examples from NMR spectroscopy in liquid solutions evidenced the flexibility and efficiency of the proposed methods. In these examples, pseudospectral methods generated pulses that achieve performance similar to that of analytical methods, and it should be noted that these ``approximated'' optimal pulses found by pseudospectral methods are always smooth. A strength of pseudospectral methods is that they provide a robust technique which is easily extendible and implementable. In addition, it has been shown empirically that these methods have exponential convergence rates \cite{trefethen_spectral_2000}, while state-of-the-art algorithms like gradient methods typically evidence linear convergence. Pseudospectral methods provide a universal tool for solving pulse design problems for dissipative quantum systems coming from different physical contexts (NMR, MRI, Quantum Optics etc). Some immediate extensions of the methods presented here include considering the optimal pulse design problem of steering a family of open or closed quantum systems with different dynamics. A concrete example is to consider a family of coupled spin systems where each one is characterized by a different relaxation rate, e.g., $\xi$ can take values from a positive interval $[\xi_1,\xi_2]$. In this case, the optimal control problem would seek to accomplish the maximum transfer, $\eta$, over all systems, namely, to maximize $\int_{\xi_1}^{\xi_2}\eta(\xi)d\xi$ subject to the system evolution as in \eqref{path1}, \eqref{path2}, or \eqref{path3}. Experimental verification of the performance of the pseudospectral pulses are also of keen interest and are currently being pursued.\\

\noindent\textbf{\small ACKNOWLEDGMENTS}\\

This work was supported by the NSF grant \#0747877.


\end{document}